# Haar Wavelets and the Origin of Gravitational Inertia

Walter J. Christensen Jr.

Cal Poly Pomona University
Physics Department
3801 W. Temple
Pomona CA 91768

wjchristense@csupomona.edu

July 28, 2007

**Abstract**

Spacetime is considered to be everywhere Minkowski except at the location where a signal wave of energy $h\nu$ interacts with the gravitational field. The conformal metric $g_{\mu\nu} \equiv f(x-vt)\eta_{\mu\nu}$ is suitably chosen to represent this interaction, where $f(x-vt)$ is a generalized wave or signal function. Parametrized and Taylor expanded at zero, the spacetime metric, $g_{\mu\nu}$, is transformed into a Haar wavelet $\tilde{\Phi}^j{}_{\mu\nu}$ having parameter width $\tau$. Applying the Haar metric to the time component of General Relativistic wave equation reduces it from a second ordered covariant differential equation to a first ordered partial differential equation. Schematically this is given by: $G\{[g^{\alpha\mu}][g_{\mu\nu}]\} \Rightarrow G\{[\tilde{\Phi}^\kappa{}_\kappa{}^{\alpha\mu}][\tilde{\Phi}^j{}_{j\mu\nu}]\} \equiv \tilde{\Phi}^j{}_{j\mu\nu,\beta} = [\alpha_j \alpha^j]_{\mu\nu} \Phi_{[j,j+1[}$. The symbol $\tilde{\Phi}^j{}_{j\mu\nu,\beta}$ represents single derivatives on the Haar metric built from step functions $\Phi_{[j,j+1[}$ and corresponding Taylor expanded coefficients. The paired coefficients $[\alpha_j \alpha^j]_{\mu\nu}$ represent energy eigenvalues (basically derivatives on the Haar spacetime metric). Because the Einstein Tensor is reduced from double covariant derivatives to a partial derivative (on the linear term of the Haar metric, but squared), the Einstein Tensor can easily be expressed in the familiar Poisson's form of: $\nabla^2 \tilde{\Phi}^1{}_{100} = -\frac{1}{\lambda^2}\Phi_{[0,1[}$, where $\Phi_{[0,1[}$ is a step function over the interval [0,1[. This expression is completely equivalent to the Newtonian expression for gravity[1]: $\nabla^2 \phi = -4\pi G\rho$, and shows the graviton wavelength to be the fundamental source for gravitational attraction. Because this wavelength is unidirectional, it strongly supports Mach's assumption that inertia arises from the total matter in the universe. Finally, since the signal metric is conformal, it can be solved for exactly, where: $g_{\mu\nu} = [1 - \sin k(x-vt)]\eta_{\mu\nu}$. Applying the signal metric to the General Relativistic gravitational equation yields an equivalent result: $G_{00} = \frac{1}{\lambda^2}\eta_{00}$.

## 1. Introduction

In 1910 Alfred Haar introduced the notion of wavelets[2]. His initial theory has been expanded recently into a wide variety of applications, but primarily it allows for the representation of various functions by a combination of step functions and wavelets over specified interval widths. The aim here is to apply techniques developed within the Haar approach to a restricted class of conformal spacetime metrics, in particular one that represents absorption of propagating signal energy into the gravitational field. Finally to write out a Poisson gravitational equation based on a Haar wavelet that shows the graviton wavelength $\lambda_g$ is the fundamental source for gravity. [By definition, the *basic wavelet* is a two-step function having a value of positive one over interval width $[0, \frac{1}{2}[$, followed by a value of negative one over the interval $[\frac{1}{2}, 1[$ . The left square bracket indicates the interval is closed and the right open, and follows the notation developed by Y. Meyer[3] on his work with Ondelettes (little waves).]

The wavelet approach to gravity begins by transforming a generalized signal metric into a Haar-metric symbolized by: $\tilde{\Phi}^j{}_{j\mu\nu}$. The Haar-metric has the property that it reduces the Einstein Tensor $G_{\mu\nu}$, from a second ordered covariant derivative equation to a first ordered partial differential equation. Schematically, the process is as follows:

$$G\{[g^{\alpha\mu}][g_{\mu\nu}]\} \Rightarrow G\{[\tilde{\Phi}^\kappa{}_\kappa{}^{\alpha\mu}][\tilde{\Phi}^j{}_{j\mu\nu}]\} \Rightarrow$$

(1.1)

$$\tilde{\Phi}^j{}_{j\mu\nu,\beta} = \{a^- a_+ \Phi_{[j,j+1[}\}_{\mu\nu} = [\alpha_j \alpha^j]_{\mu\nu} \Phi_{[j,j+1[}$$

It will be shown that $a^- a_+$ are partial derivative operators acting on the Haar-metric and play a similar role, as raising and lowering operators do in quantum mechanics, but for gravity. This can be understood by examining the Taylor expanded signal metric and its inverse, where higher orders in the interval width $\tau$ have been dropped because $\tau$ is infinitely small. Thus:

$$g_{\mu\nu} = [\eta_{\mu\nu} + f'(0)\tau \cdot \eta_{\mu\nu}]$$ (1.2)

$$g^{\mu\nu} = [\eta^{\mu\nu} - f'(0)\tau \cdot \eta_{\mu\nu}]$$ (1.3)

Since the second term in both metrics represents a graviton formed from signal energy, and that the Einstein tensor is directly related to the Energy Momentum Tensor, derivatives on the Haar metric will produce positive and negative graviton energies.

## 2. Gravitational Model

For simplicity sake, the model presented here assumes spacetime is everywhere Minkowski, except at the location where a signal wave $f(x-vt)$ interacts with the gravitational field. Since propagating waves possess energy, and in turn affect the curvature of spacetime, the metric:

$$g_{\mu\nu} \equiv f(x-vt)\eta_{\mu\nu} \tag{2.1}$$

suitably represents the absorption of signal wave-energy by flat spacetime. Because Haar wavelets are fundamentally represented as an outer product of 1-forms[4], and that the goal here is to represent the signal metric $f(x-vt)\eta_{\mu\nu}$ as a Haar wavelet, the starting point is to show the Minkowski metric can also be represented as an outer product of 1-forms[5]:

$$\eta_{\mu\nu} \equiv (\tilde{p} \otimes \tilde{q})(e_\mu, e_\nu) \equiv \begin{bmatrix} -1 & 0 & 0 & 0 \\ 0 & 1 & 0 & 0 \\ 0 & 0 & 1 & 0 \\ 0 & 0 & 0 & 1 \end{bmatrix} \tag{2.2}$$

Next, the signal function $f(x-vt)$ is parameterized and expanded into a Taylor series at $\tau = 0$[6], such that:

$$g_{\mu\nu} = f(\tau)\eta_{\mu\nu} = f(0)\eta_{\mu\nu} + \frac{f'(0)\tau}{1!}\eta_{\mu\nu} + \frac{f''(0)\tau^2}{2!}\eta_{\mu\nu} + \ldots \tag{2.3}$$

For reasons of compact notation, the Taylor coefficients and interval widths $\tau$ are defined in the following manner:

$$\bar{a} = (a_0, a_1, a_2, \ldots, a_j) = f(0), \frac{f'(0)\tau}{1!}, \frac{f''(0)\tau^2}{2!}, \ldots \tag{2.4}$$

To keep the background geometry of spacetime flat[7], the coefficient $a_0$ is defined to be unity. It is also important to note that each of the coefficients $\{a_j\}$ contain powers of the interval width $\tau$. In wavelet theory interval width determines how much (or little) information is required to accurately describe the physical system or function at hand; this is called the sample. With Haar wavelet theory one typically seeks a minimal set of interval widths. Usually this is accomplished through an algorithm that absorbs smaller widths into larger widths[8], yet keeps the integrity of the original function or physical system under consideration. In abbreviated notation, the expanded signal metric can be rewritten as:

$$g_{\mu\nu} = f(\tau)\eta_{\mu\nu} = \sum_{j=0}^{n} a_j \eta_{\mu\nu} \tag{2.5}$$

It will be shown in section 4 that coefficients $a_j$ always form pairs, each with their own partial derivatives, such that:

$$\partial_\mu \tau = \begin{cases} \pm v \\ 1 \end{cases} \tag{2.6}$$

where $\pm v$ is the velocity of the traveling signal wave. Because derivative coefficients are paired, they typically will form an expression for classical and special relativistic energy densities: $\frac{1}{2}\rho v^2, \rho c^2$. The conclusion is that Haar wavelets carry signal energies that are absorbed into Minkowski spacetime to produce graviton particles. It will be shown for the time component of the transformed Einstein Tensor $G_{00}$, all velocity-squared terms cancel out, leaving a constant term related to the graviton wavelength $\lambda_g$, such that:

$$G_{00} = -\kappa^2 = -\frac{1}{\lambda_g^2} \tag{2.7}$$

[Note: By expanding at zero an observer can be interpreted as moving along with the propagating wave, where $x \cong vt$. This is important in two regards: First the observer is moving in an inertial frame with constant velocity v; secondly observation and measurement occurs at a point in spacetime, and therefore yields something more fundamental about nature. As Albert Einstein pointed out: "Firstly we see that at any point filled with matter, there exists a preferred state of motion, namely that of the substance at that point considered…The rigid body is only achieved approximately in nature, not even with desired approximation."[9]. By Taylor expanding the signal metric, interval width $\tau$ naturally falls out and allows interval width to be defined in a new way (see below). Since $\tau$ is infinitely small, all higher orders of Taylor expanded terms can be dropped from the expansion (this will be explained in more detail later), leaving two important terms to represent the signal metric, namely:

$$\eta_{\mu\nu} + f'(0)\tau \cdot \eta_{\mu\nu} \tag{2.8}$$

The first term, $\eta_{\mu\nu}$, is simply the background kinematic metric having zero ground-state energy. The second term $f'(0)\tau$ contains the signal energy absorbed into the flat geometry of spacetime, hence represents a graviton.]

The next step in transforming the signal metric into a Haar wavelet, is to define a step function such that:

$$\Phi_{[j,j+1[}(r) := \begin{cases} +1 & \text{if } j \leq r < j+1 \\ 0 & \text{otherwise} \end{cases} \qquad (2.9)$$

where $j = 0,1,2,3...$, and where r is an element of the real number.

Continuing on, a new function, $\tilde{\Phi}_{j\mu\nu}$, is constructed from the coefficients $a_j$ and step function $\Phi_{[j,j+1[}(r)$, such that:

$$\tilde{\Phi}_{j\mu\nu} \equiv \left[ \alpha_j \cdot \Phi_{[j,j+1[}(r) \right]_{\mu\nu} \equiv \begin{cases} +\alpha_j & \text{if } \mu = \nu = 1,2,3 \\ -\alpha_j & \text{if } \mu = \nu = 0 \\ 0 & \text{if } \mu \neq \nu \end{cases} \qquad (2.10)$$

[If the indices $\mu = \nu = 0$ are contravariant, then the sign changes to: $+\alpha_j$. Using Einstein's summation notation, the signal metric can be compactly represented by a series of step functions and coefficients, where:

$$g_{\mu\nu} \equiv \tilde{\Phi}^j{}_{j\mu\nu} \qquad (2.11)$$

In summed out form the signal metric appears as:

$$\tilde{\Phi}^j{}_{j\mu\nu} = \tilde{\Phi}^0{}_{0\mu\nu} + \tilde{\Phi}^1{}_{1\mu\nu} + \tilde{\Phi}^2{}_{2\mu\nu} + ... =$$
$$\left( \alpha_0 \Phi_{[0,1[} + \alpha_1 \Phi_{[1,2[} + \alpha_2 \Phi_{[2,3[} + ... \right)_{\mu\nu} \qquad (2.12)$$

Because each of the step functions: $\Phi_{[j,k+1[}$ all have the same value over their respective interval, the step function can be factored out from the summation (see equation 1.5 in its Minkowski form to see this) and its interval made as large or small as needed. It is standard to define Haar wavelets over the interval [0,1[. Thus the preceding equation can be rewritten as:

$$\tilde{\Phi}^j{}_{j\mu\nu} = (\alpha_0 + \alpha_1 + \alpha_2 + \cdots \alpha_{k+1})(\Phi_{[0,1[})_{\mu\nu} \qquad (2.13)$$

At this point, a necessary condition is imposed on the interval width $\tau$, such that for all x and t, let $(x - vt)$ be positive, that is $\tau > 0$. Then for all x and t, the coefficients $a_j$ can be organized into positive and negative groups, such that:

$$\tilde{\Phi}^j{}_{j\mu\nu} = A \cdot \Phi_{[0,\frac{1}{2}[} + B \cdot \Phi_{[\frac{1}{2},1[} \qquad (2.14)$$

where for each x and t, A is a maximum positive number of summed positive real coefficients $\alpha_j$. Likewise B is the sum of all negative coefficients $\alpha_k$. The existence of A and B can be argued by noting that a generalized wave function can be represented by trigonometric functions, and that a Taylor expansion of say a sine or cosine function will produce constants of derivatives on that function evaluated at the expansion point and produce negative or positive terms in the infinite sum. Thus, for each point (x, t), a Haar wavelet can be constructed from equation (2.14), such that:

$$\frac{\tilde{\Phi}^j{}_{j\mu\nu}}{C} = \begin{cases} 1 & \text{if } 0 \leq m < \frac{1}{2} \text{ and } C = A \\ -1 & \text{if } \frac{1}{2} \leq m < 1 \text{ and } C = B \end{cases} \tag{2.15}$$

The signal metric can therefore be represented by a Haar wavelet:

$$g_{\mu\nu} = C \frac{\tilde{\Phi}^j{}_{j\mu\nu}}{C} \tag{2.16}$$

## 3. Closer Examination of the Haar metric

One important design feature of Haar-Einstein equation is that the metric and its inverse necessarily form contracted Minkowski metric pairs, leaving behind paired coefficients--each with their own partial derivative. Since they often take on the value of the velocity of the signal wave, and that Haar-Einstein equation is directly related to the Energy momentum Tensor, the coefficients reveal their relation to signal energy and do so simply because the Einstein Tensor produces velocity squared terms. This result, and that the initial condition states spacetime absorbs a quantum of energy $h\nu$, allows for the Haar-Einstein gravitational equation to be solved exactly (this will be shown in subsequent sections).

At this point a closer examination of the properties of the Haar metric is carried out. It is necessary, therefore, to express the inverse metric $g^{\mu\nu}$, in the form of a Haar wavelet. This is readily accomplished by asking what tensor contracts with the signal metric $f(x-vt)\eta_{\mu\nu}$, and yields the Dirac delta function? The immediate conclusion is:

$$g^{\mu\nu} \equiv \frac{\eta^{\mu\nu}}{f(x-vt)} \tag{3.1}$$

As a check, the signal metric is contracted with its signal inverse:

$$g_{\alpha\nu} g^{\mu\nu} \equiv f(x-vt)\eta_{\alpha\nu} \frac{\eta^{\mu\nu}}{f(x-vt)} = \delta_\alpha^\mu \tag{3.2}$$

As before, the inverse signal metric is Taylor expanded to yield its Haar wavelet form:

$$g^{\mu\nu} \equiv \Phi_\kappa{}^{\kappa\mu\nu} = \sum_{\kappa=0}^{K} a^\kappa \eta^{\mu\nu} \qquad (3.3)$$

Keeping only first order terms of the expanded metrics, and approximating square powers of $\tau$ to zero, the Haar metric and inverse are contracted to yield:

$$\eta^{\alpha\mu}[1 - f'(0)\tau] \cdot \eta_{\mu\nu}[1 + f'(0)\tau] =$$

$$[1 + f'(0)^2 \tau^2]\Phi_{[0,1[}\delta_\nu^\alpha = 1 = \Phi_{[0,1[}(r) \qquad (3.4)$$

By producing a step function from the contraction of two Haar wavelets metrics, completes the development of the Haar metric. However, it is noteworthy to point out this contraction produces a series of paired coefficients $\{a_j\}$ and $\{a^\kappa\}$, which sum to unity. This result is similar to the normalization condition imposed on the coefficients during the development of second quantization of the Dirac field[10], where:

$$\frac{1}{n!}\sum |c(\alpha_1, \alpha_2, ..., \alpha_n)|^2 = 1 \qquad (3.5)$$

## 4. Coefficients of the Haar Metric

The spacetime signal metric expanded to 1$^{st}$ order is given by:

$$g_{\mu\nu} = \eta_{\mu\nu} + f'(0)\tau \cdot \eta_{\mu\nu} \qquad (4.1)$$

Comparing the preceding equation with: $g_{\mu\nu} \Rightarrow \tilde{\Phi}^j{}_{j\mu\nu} = \sum_{j=0}^{j} \alpha_j \eta_{\mu\nu}$, we see:

$$\alpha_0 = 1, \ \alpha_1 = f'(0)\tau \qquad (4.2)$$

And so:

$$g_{\mu\nu} = [\eta_{\mu\nu} + \alpha_1 \eta_{\mu\nu}] \qquad (4.3)$$

Note, to first order that the Haar metric $\tilde{\Phi}^j{}_{j\mu\nu}$ is equivalent to Einstein's weak field metric for gravitational waves[11,12,13]

$$g_{\mu\nu} \equiv -\eta_{\mu\nu} + h_{\mu\nu} \qquad (4.4)$$

Likewise the contravariant Haar metric is given by:

$$g^{\mu\nu} - \alpha^1 \eta^{\mu\nu} \tag{4.5}$$

It is apparent that:

$$\alpha^1 = \alpha_1 \tag{4.6}$$

**5. Transforming the time component of Einstein Tensor into Wavelet form**

The General relativistic gravitational wave equation, developed by Einstein and to some degree by Hilbert[14], is given by:

$$G_{00} = R_{00} - \frac{1}{2} g_{00} R = -8\pi G T_{00} \tag{5.1}$$

Since all double derivatives on the expanded Haar metric are immediately zero [see equation (5.4)], the Ricci Curvature tensor reduces to:

$$R_{00} \equiv \Gamma^{\lambda}_{0\lambda,0} - \Gamma^{\lambda}_{00,\lambda} \tag{5.2}$$

Double derivatives on the Christoffel symbol vanish and so are not included here. The first term $\Gamma^{\lambda}_{0\lambda,0}$ is examined:

$$\Gamma^{\lambda}_{0\lambda} = \frac{1}{2} g^{\lambda\nu} \{g_{\lambda\nu,0} + g_{0\nu,\lambda} - g_{\lambda 0,\nu}\} \tag{5.3}$$

Its derivative becomes:

$$\Gamma^{\lambda}_{0\lambda,0} = \frac{1}{2} \{g^{\lambda\nu} g_{\lambda\nu,00} + g^{\lambda\nu} g_{0\nu,\lambda 0} - g^{\lambda\nu} g_{\lambda 0,\nu 0}\} + \{g^{\lambda\nu}{}_{,0} g_{\lambda\nu,0} + g^{\lambda\nu}{}_{,0} g_{0\nu,\lambda} - g^{\lambda\nu}{}_{,0} g_{\lambda 0,\nu}\} \tag{5.4}$$

Because $g_{\lambda\nu,\alpha\beta} = 0$ for all α and β, the previous expression reduces to:

$$\Gamma^{\lambda}_{0\lambda,0} = \frac{1}{2} g^{\lambda\nu}{}_{,0} g_{\lambda\nu,0} \tag{5.5}$$

In a similar manner, the second Christoffel symbol $\Gamma^{\lambda}_{00,\lambda}$ is computed to be:

$$\Gamma^{\lambda}_{00} = \frac{1}{2} \{g^{\lambda\nu} g_{0\nu,0} + g^{\lambda\nu} g_{0\nu,0} - g^{\lambda\nu} g_{00,\nu}\} = g^{\lambda\nu} g_{0\nu,0} - \frac{1}{2} g^{\lambda\nu} g_{00,\nu} \tag{5.6}$$

Its derivative becomes:

$$\Gamma^{\lambda}_{00,\lambda} = g^{\lambda v}{}_{,\lambda} g_{0v,0} - \frac{1}{2} g^{\lambda v}{}_{,\lambda} g_{00,v} \qquad (5.7)$$

$$R_{00} = \frac{1}{2} g^{\lambda v}{}_{,0} g_{\lambda v,0} - g^{\lambda v}{}_{,\lambda} g_{0v,0} + \frac{1}{2} g^{\lambda v}{}_{,\lambda} g_{00,v} \qquad (5.8)$$

To transform the time component of Riemannian Curvature tensor into the related time component of the Haar Einstein gravitational equation, the Haar metric and its inverse:

$$g_{\mu v} \Rightarrow \eta_{\mu v} + \alpha_1 \cdot \eta_{\mu v} \qquad (5.9)$$

$$g^{\mu v} \Rightarrow \eta_{\mu v} - \alpha^1 \eta^{\mu v} \qquad (5.10)$$

is substituted into the wave equation. From equation (5.8) each term of $R_{00}$ is transformed by. We begin with the first term $\frac{1}{2} g^{\lambda v}{}_{,0} g_{\lambda v,0}$, where:

$$g^{\lambda v}{}_{,0} = [\eta^{\lambda v} - \alpha^1 \eta^{\lambda v}]_{,0} = -\alpha^1{}_{,0} \eta^{\lambda v} \qquad (5.11)$$

Likewise:

$$g_{\lambda v,0} = [\eta_{\lambda v} + \alpha_1 \eta_{\lambda v}]_{,0} = \alpha_{1,0} \eta_{\lambda v} \qquad (5.12)$$

Given that $\eta^{\lambda v} \eta_{\lambda v} = 4$, the first term of $R_{00}$ is transformed into:

$$\frac{1}{2} g^{\lambda v}{}_{,0} g_{\lambda v,0} = -\frac{1}{2} [\alpha^1{}_{,0} \alpha_{1,0}] \eta^{\lambda v} \eta_{\lambda v} =$$

$$- 2[a^1{}_{,0} a_{1,0}] \qquad (5.13)$$

Continuing on, the second term of $R_{00}$, given by: $[-g^{\lambda v}{}_{,\lambda} g_{0v,0}]$, transforms as follows:

$$g^{\lambda v}{}_{,\lambda} = -\alpha^1{}_{,\lambda} \eta^{\lambda v} \qquad (5.14)$$

$$g_{0v,0} = \alpha_{1,0} \eta_{0v} \qquad (5.15)$$

The second term simplifies to:

$$-g^{\lambda\nu}{}_{,\lambda} g_{0\nu,0} = [\alpha^1{}_{,0}\alpha_{1,0}]\eta^{\lambda\nu}\eta_{0\nu} =$$

$$[\alpha^1{}_{,0}\alpha_{1,0}] \tag{5.16}$$

The last term of $\frac{1}{2} g^{\lambda\nu}{}_{,\lambda} g_{00,\nu}$ is transforms into:

$$g^{\lambda\nu}{}_{,\lambda} = -\alpha^1{}_{,\lambda}\eta^{\lambda\nu} \tag{5.17}$$

$$g_{00,\nu} = \alpha_{1,\nu}\eta_{00} \tag{5.18}$$

Therefore the third term becomes:

$$\frac{1}{2} g^{\lambda\nu}{}_{,\lambda} g_{00,\nu} = -\frac{1}{2}[\alpha_{1,\nu}\alpha^1{}_{,\lambda}]\eta_{00}\eta^{\lambda\nu} =$$

$$-\frac{1}{2}\{[\alpha_{1,0}\alpha^1{}_{,0}]\eta_{00}\eta^{00} + [\alpha_{1,1}\alpha^1{}_{,1}]\eta_{00}\eta^{11} + [\alpha_{1,2}\alpha^1{}_{,2}]\eta_{00}\eta^{22} + [\alpha_{1,3}\alpha^1{}_{,3}]\eta_{00}\eta^{33}\} \tag{5.19}$$

Since $\alpha_j = f'(0)(x - vt)$, derivatives 2 and 3 on $\alpha_j$ vanish. Thus the third term of $R_{00}$ becomes:

$$\frac{1}{2} g^{\lambda\nu}{}_{,\lambda} g_{00,\nu} = -\frac{1}{2}\{[\alpha_{1,0}a^1{}_{,0}]\eta_{00}\eta^{00} + [\alpha_{1,1}\alpha^1{}_{,1}]\eta_{00}\eta^{11}\} =$$

$$-\frac{1}{2}[\alpha_{1,0}\alpha^1{}_{,0}] + \frac{1}{2}[\alpha_{1,1}\alpha^1{}_{,1}] \tag{5.20}$$

Combining all terms for $R_{00}$ yields:

$$R_{00} = -2[\alpha^1{}_{,0}\alpha_{1,0}] + [\alpha^1{}_{,0}\alpha_{1,0}] - \frac{1}{2}[\alpha_{1,0}\alpha^1{}_{,0}] + \frac{1}{2}[\alpha_{1,1}\alpha^1{}_{,1}] \tag{5.21}$$

$R_{00}$ can further be reduced from the conditions:

$$\alpha^1{}_{,0} = \alpha_{1,0} \tag{5.22}$$

and

$$\alpha^{1,1} = \alpha_{1,1} \tag{5.23}$$

Thus

$$R_{00} = -2(\alpha_{1,0})^2 + (\alpha_{1,0})^2 - \frac{1}{2}(\alpha_{1,0})^2 + \frac{1}{2}(\alpha_{1,1})^2 =$$

$$-\frac{3}{2}(\alpha_{1,0})^2 + \frac{1}{2}(\alpha_{1,1})^2$$

(5.24)

The next term of $G_{00}$ to evaluate is the scalar curvature tensor R. Because of the linear nature of the signal metric, again double derivates of the Christoffel symbol vanish and so are not included here:

$$R \equiv g^{\mu\kappa} R_{\mu\kappa} = \{g^{\mu\kappa} \Gamma^{\lambda}_{\mu\lambda,\kappa} - g^{\mu\kappa} \Gamma^{\lambda}_{\mu\kappa,\lambda}\}$$

(5.25)

The first term $g^{\mu\kappa}\Gamma^{\lambda}_{\mu\lambda,\kappa}$ expands to:

$$g^{\mu\kappa}\Gamma^{\lambda}_{\mu\lambda,\kappa} = \frac{1}{2}g^{\mu\kappa}g^{\lambda\nu}{}_{,\kappa}\{g_{\lambda\nu,\mu} + g_{\mu\nu,\lambda} - g_{\lambda\mu,\nu}\} + \frac{1}{2}g^{\mu\kappa}g^{\lambda\nu}\{g_{\lambda\nu,\mu\kappa} + g_{\mu\nu,\lambda\kappa} - g_{\lambda\mu,\nu\kappa}\}$$

(5.26)

Since all double derivative terms of the metric are zero, the previous expression reduces to:

$$g^{\mu\kappa}\Gamma^{\lambda}_{\mu\lambda,\kappa} = \frac{1}{2}g^{\lambda\nu,\mu}g_{\lambda\nu,\mu}$$

(5.27)

The second Christoffel term $g^{\mu\kappa}\Gamma^{\lambda}_{\mu\kappa,\lambda}$ is evaluated in a similar manner:

$$g^{\mu\kappa}\Gamma^{\lambda}_{\mu\kappa,\lambda} = \frac{1}{2}\{g^{\mu\kappa}g^{\lambda\nu}{}_{,\lambda}g_{\kappa\nu,\mu} + g^{\mu\kappa}g^{\lambda\nu}{}_{,\lambda}g_{\mu\nu,\kappa} - g^{\mu\kappa}g^{\lambda\nu}{}_{,\lambda}g_{\kappa\mu,\nu}\}$$

(5.28)

$$g^{\mu\kappa}\Gamma^{\lambda}_{\mu\kappa,\lambda} = \left\{g^{\lambda\nu}{}_{,\lambda}g_{\kappa\nu}{}^{,\kappa} - \frac{1}{2}g^{\mu\kappa}g^{\lambda\nu}{}_{,\lambda}g_{\kappa\mu,\nu}\right\}$$

(5.29)

Thus:

$$R = \frac{1}{2}g^{\lambda\nu,\mu}g_{\lambda\nu,\mu} - g^{\lambda\nu}{}_{,\lambda}g_{\kappa\nu}{}^{,\kappa} + \frac{1}{2}g^{\mu\kappa}g^{\lambda\nu}{}_{,\lambda}g_{\kappa\mu,\nu}$$

(5.30)

R is now transformed into Haar metric form beginning with the first term $\frac{1}{2}g^{\lambda\nu,\mu}g_{\lambda\nu,\mu}$:

$$g^{\lambda\nu,\mu} = -\alpha^{1,\mu}\eta^{\lambda\nu} \tag{5.31}$$

$$g_{\lambda\nu,\mu} = \alpha_{1,\mu}\eta_{\lambda\nu} \tag{5.32}$$

and so the first term of R becomes:

$$-\frac{1}{2}[\alpha^{1,\mu}\alpha_{1,\mu}]\eta^{\lambda\nu}\eta_{\lambda\nu} = -\frac{1}{2}[\alpha^{1,0}\alpha_{1,0}]\eta^{\lambda\nu}\eta_{\lambda\nu} - \frac{1}{2}[\alpha^{1,1}\alpha_{1,1}]\eta^{\lambda\nu}\eta_{\lambda\nu} =$$

$$-2[\alpha^{1,0}\alpha_{1,0}] - 2[\alpha^{1,1}\alpha_{1,1}] \tag{5.33}$$

[Careful, covariant time derivatives have the extra negative sign.]

Continuing on with the transformation of R, the second term $[-g^{\lambda\nu}{}_{,\lambda}g_{\kappa\nu}{}^{,\kappa}]$ becomes:

$$g^{\lambda\nu}{}_{,\lambda} = -\alpha^{1}{}_{,\lambda}\eta^{\lambda\nu} \tag{5.34}$$

$$g_{\kappa\nu}{}^{,\kappa} = \alpha_{1,}{}^{\kappa}\eta_{\kappa\nu} \tag{5.35}$$

Thus $-g^{\lambda\nu}{}_{,\lambda}g_{\kappa\nu}{}^{,\kappa}$ transforms into:

$$[\alpha^{1}{}_{,\lambda}\alpha_{1}{}^{\kappa}]\eta^{\lambda\nu}\eta_{\kappa\nu} = [\alpha^{1}{}_{,0}\alpha_{1}{}^{0}]\eta^{00}\eta_{00} + [\alpha^{1}{}_{,1}\alpha_{1}{}^{1}]\eta^{11}\eta_{11} =$$

$$[\alpha^{1}{}_{,0}\alpha_{1}{}^{0}] + [\alpha^{1}{}_{,1}\alpha_{1}{}^{1}] \tag{5.36}$$

The third and final term, $[\frac{1}{2}g^{\mu\kappa}g^{\lambda\nu}{}_{,\lambda}g_{\kappa\mu,\nu}]$, has the extra consideration of:

$$g^{\mu\kappa} = \eta^{\mu\kappa} - \alpha^{1}\eta^{\mu\kappa} \tag{5.37}$$

However $\alpha^{1}\eta^{\mu\kappa}$ maybe dropped from $[\eta^{\mu\kappa} - \alpha^{1}\eta^{\mu\kappa}]$, because $\alpha^{1} = f'(0)\tau$, but the Taylor expansion imposes the condition that: $\tau << 1$. And so:

$$g^{\mu\kappa} = \eta^{\mu\kappa} \tag{5.38}$$

$$g^{\lambda\nu}{}_{,\lambda} = -\alpha^{1}{}_{,\lambda}\eta^{\lambda\nu} \tag{5.39}$$

$$g_{\kappa\mu,\nu} = \alpha_{1,\nu}\eta_{\kappa\mu} \tag{5.40}$$

Therefore, third term: $\frac{1}{2} g^{\mu\kappa} g^{\lambda\nu}{}_{,\lambda} g_{\kappa\mu,\nu}$ transforms to:

$$-\frac{1}{2}\alpha^1{}_{,\lambda}\alpha_{1,\nu}\eta^{\lambda\nu}\eta_{\kappa\mu}\eta^{\mu\kappa} \qquad (5.41)$$

Given that $\eta^{\mu\kappa}\eta_{\kappa\mu} = 4$ and $\eta^{00} = \eta_{00} = -1$, the third term becomes:

$$-2\alpha^1{}_{,\lambda}\alpha_{1,\nu}\eta^{\lambda\nu} = -2\alpha^1{}_{,0}\alpha_{1,0}\eta^{00} - 2\alpha^1{}_{,1}\alpha_{1,1}\eta^{11} =$$

$$2\alpha^1{}_{,0}\alpha_{1,0} - 2\alpha^1{}_{,1}\alpha_{1,1} \qquad (5.42)$$

Therefore, R becomes:

$$R = -2[a^{1,0}a_{1,0}] - 2[a^{1,1}a_{1,1}] + [a^1{}_{,0}a_1{}^0] + [a^1{}_{,1}a_1{}^1] + 2a^1{}_{,0}a_{1,0} - 2a^1{}_{,1}a_{1,1}] \qquad (5.43)$$

As before R can be reduced to:

$$R = 3(a_{1,0})^2 - 3(a_{1,1})^2 \qquad (5.44)$$

And since: $-\frac{1}{2}g^{00}R \cong -\frac{1}{2}\eta^{00}R = \frac{1}{2}R$, then:

$$-\frac{1}{2}g^{00}R = +\frac{3}{2}(a_{1,0})^2 - \frac{3}{2}(a_{1,1})^2 \qquad (5.45)$$

Putting together the Haar-Einstein time component of the tensor, $G_{00}$, yields:

$$-\frac{3}{2}(a_{1,0})^2 + \frac{1}{2}(a_{1,1})^2 + \frac{3}{2}(a_{1,0})^2 - \frac{3}{2}(a_{1,1})^2 \qquad (5.46)$$

Therefore:

$$G_{00} = -(a_{1,1})^2 \qquad (5.47)$$

where:

$$a_{1,1} = f'(0)(x-vt)_{,x} = f'(0) \qquad (5.48)$$

Therefore the time component of Einstein's field equations becomes:

$$G_{00} = -f'(0)^2 \qquad (5.49)$$

Can the constant $f'(0)^2$ be determined? The answer will be yes. To show this the Einstein tensor result is equated to the energy momentum tensor, such that:

$$f'(0)^2 = \frac{8\pi G}{c^4} T_{00} \qquad (5.50)$$

Note that the left hand side of the preceding equation is unitless, whereas the right hand side has units of $meters^{-2}$. This imbalance can be rectified by redefining the interval to be: $\tau \equiv \kappa(x - vt)$. The signal metric becomes unitless, such that:

$$g_{\mu\nu} \cong \eta_{\mu\nu} + \kappa \cdot f'(0)(x - vt)\eta_{\mu\nu} \qquad (5.51)$$

where $\kappa = \frac{1}{\lambda_g}$ (where $\lambda_g$ is the wavelength of a graviton). The time component of the Einstein tensor acting on adjusted signal metric produces the constant result:

$$G_{00} = (a_{1,1})^2 = \kappa^2 \cdot f'(0)^2 \qquad (5.52)$$

For reasons that will soon become apparent, the Einstein Tensor $G_{00}$ is now multiplied by: $\frac{1}{\kappa^4} = \lambda_g^{\,4}$ so that:

$$\frac{1}{\kappa^4} G_{00} = -\frac{1}{\kappa^2} f'(0)^2 = -\frac{8\pi G}{\kappa^4 c^4} T_{00} \qquad (5.53)$$

[Historically, the wave number $\kappa$ is closely connected to many important quantum mechanical results, and so it will be here].

Since the wavelength of a graviton is known up to several orders of accuracy, this leaves two unknowns $\{f'(0), T_{00}\}$, but only one equation to solve for these unknowns. However, a second equation exists from the initial condition of signal energy $\hbar c \kappa$ and $T_{00}$. The total energy is known from the initial condition, but what volume of spacetime contains energy $\hbar c \kappa$. Can we simply assume?:

$$T_{00} = \frac{\hbar c \kappa}{m^3} \qquad (5.54)$$

Certainly this cannot be correct. For if the volume were cut in half, it would contain the same total energy, yet the density should not go up eight fold. An observer moving along with the propagating signal wave would not measure either of these energy densities. Somehow the preceding expression (5.54) must be scaled appropriately. Since the graviton wavelength is incredibly long:

$$\lambda_g = \frac{1}{\kappa} = \frac{c}{\nu} = 1.105 \times 10^{23} \, meters \tag{5.55}$$

and the energy is spread over that length, it is reasonable to assume the unitless scaling factor $\tilde{\lambda}_g$ should arise from the wavelength. The choice made for the scaling factor that works is:

$$volume \equiv \frac{m^3}{\tilde{\lambda}_g^{\,2}} \tag{5.56}$$

For reasons of notational compactness, let the tilde be dropped from the lambda and kappa symbols. Since the graviton wavelength is so incredibly long, the scaling factor quantizes volume. Writing lambda in terms of its wave number yields:

$$-\frac{1}{\kappa^2} f'(0)^2 = -\frac{8\pi G}{\kappa^6 c^4} \frac{\hbar c \kappa}{m^3} \tag{5.57}$$

Solving for the unknown constant yields:

$$f'(0)^2 = \frac{8\pi G \hbar}{\kappa^3 c^3} = \frac{8\pi \lambda^3 G \hbar}{c^3} \tag{5.58}$$

The wavelength, $\lambda = \frac{1}{\kappa}$, of a graviton[15] can be derived from the mass of the graviton and the de Broglie graviton wave-frequency:

$$\nu_g = \frac{mc^2}{2\pi\hbar} = 2.7 \times 10^{-15} \, \sec^{-1} \tag{5.59}$$

The wavelength for a graviton becomes:

$$\lambda_g = \frac{1}{\kappa} = \frac{c}{\nu} = 1.105 \times 10^{23} \, meters \tag{5.60}$$

Solving for $f'(0)^2$, yields:

$$f'(0)^2 = \frac{8\pi(\lambda)^3 G\hbar}{c^3} = 8.8 \tag{5.61}$$

However, this value depends on the graviton mass, which can be off by three or more orders of magnitude[16,17]. It is therefore assumed that $f'(0)^2$ has a value of unity. If so, then a fundamental equation for gravity based on a single graviton can be expressed as:

$$[\lambda^4 G_{00}] = \lambda^2 \eta_{00} \tag{5.62}$$

This can also be expressed more simply as:

$$G_{00} = \kappa^2 \eta_{00} \tag{5.63}$$

where $\kappa = \dfrac{1}{\lambda_g}$.

Finally, in Haar-Einstein notation, equation (5.62) becomes:

$$\tilde{\Phi}^1{}_{100,x} = -\frac{1}{\lambda^2} \Phi_{[0,1[} \tag{5.64}$$

Since there are no y and z components in the signal metric equation and that $G_{00} = (a_{1,1})^2 = \kappa^2$, the preceding equation can be written as:

$$G_{00} \Rightarrow \left(i \cdot \nabla \tilde{\Phi}^1{}_{100}\right)^2 = -\frac{1}{\lambda^2} \tilde{\Phi}_{[0,1[} \tag{5.65}$$

However, since $G_{00} \Rightarrow -(a_{1,1})^2 = -\left(\kappa \cdot (x-vt)_{,x}\right)^2 = -\kappa^2$, if the wavelet is redefined to be:

$$\tilde{\Phi}^1{}_{100} \equiv -\frac{1}{2}[\kappa(x-vt)]^2 \Phi_{[0,1[} \tag{5.66}$$

then equation (4.63) can be put into the very familiar Poisson form of:

$$G_{00} \Rightarrow \nabla^2 \tilde{\Phi}^1{}_{100} = -\frac{1}{\lambda^2} \Phi_{[0,1[} \tag{5.67}$$

Since this form is equivalent to the gravitational expression of:

$$\nabla^2 \phi = -4\pi\rho \tag{5.68}$$

it is immediately realized that $\lambda_g$ must be the source for gravitational attraction. Furthermore, since this attraction is unidirectional, or in the direction of the traveling signal wave, it strongly supports Mach's assumption that inertia arises from the total matter of the universe. As Albert Einstein stated, "the $g_{\mu\nu}$ field shall be completely determined by matter."[18]. Equation (4.60) also shows gravity radial dependence results from interval $\tau \to 0$ such that the force of gravity is proportional to:

$$\frac{1}{r^2} \tag{5.69}$$

Because the Taylor coefficients were determined to be:

$$f(0) = f'(0)^2 = 1$$

this condition, and that the metric is conformal, allows for the exact solution of the signal metric. This conformal function is given by:

$$f(x - vt) = 1 - \sin \kappa (x - vt) \tag{5.70}$$

$$f(x - vt) = 1 - \sin \kappa (x - vt) \tag{5.71}$$

Thus the signal metric becomes:

$$g_{\mu\nu} = [1 - \sin \kappa (x - vt)] \eta_{\mu\nu} \tag{5.72}$$

Its inverse is given by:

$$g^{\mu\nu} = \frac{1}{[1 - \sin \kappa (x - vt)]} \eta^{\mu\nu} \tag{5.73}$$

For these metrics to give the same answer as the Haar metric, they must be evaluated at $\tau = (x - vt) \Rightarrow 0$. Since this entails the derivative of the metric with respect to time t or position x, a representative calculation is shown for the time component of the inverse metric:

$$g^{00}{}_{,0} = \left(\frac{1}{[1 - \sin \kappa (x - vt)]}\right)_{,0} \eta^{00} = -\left(\frac{1}{[1 - \sin \kappa (x - vt)]}\right)_{,0} \tag{5.74}$$

Careful the time derivative has an extra minus sign associated with it. Thus:

$$g^{00}{}_{,0} = -\frac{\kappa v \cos \kappa (x - vt)}{[1 - \sin \kappa (x - vt)]^2} \tag{5.75}$$

However the metrics are always paired in the gravitational equation with its inverse. For example: $R_{00} = \frac{1}{2} g^{\lambda\nu}{}_{,0} g_{\lambda\nu,0} - g^{\lambda\nu}{}_{,\lambda} g_{0\nu,0} + \frac{1}{2} g^{\lambda\nu}{}_{,\lambda} g_{00,\nu}$. Examining the first term leads to:

$$\frac{1}{2} g^{\lambda\nu}{}_{,0} g_{\lambda\nu,0} = \frac{1}{2} \eta^{\lambda\nu} \left( \frac{1}{1 - \sin\kappa(x - vt)} \right)_{,0} \eta_{\lambda\nu} [1 - \sin\kappa(x - vt)]_{,0} =$$

(5.76)

$$-2 \frac{\kappa^2 v^2 \cos^2[\kappa(x - vt)]}{[1 - \sin\kappa(x - vt)]^2}$$

Next the denominator is multiplied by its conjugate:

$$-2 \frac{\kappa^2 v^2 \cos^2[\kappa(x - vt)]}{[1 - \sin\kappa(x - vt)]^2} \frac{[1 + \sin\kappa(x - vt)]^2}{[1 + \sin\kappa(x - vt)]^2} =$$

(5.77)

$$-2 \frac{\kappa^2 v^2 \cos^2[\kappa(x - vt)][1 + \sin\kappa(x - vt)]^2}{[1 - \sin^2\kappa(x - vt)]^2} =$$

$$-2 \frac{\kappa^2 v^2 \cos^2[\kappa(x - vt)][1 + \sin\kappa(x - vt)]^2}{\cos^2[\kappa(x - vt)]} =$$

(5.78)

$$-2\kappa^2 v^2 [1 + \sin\kappa(x - vt)]^2$$

However, to match up results with the Haar metric, the sine function must also be evaluated infinitely close to zero. Since $\kappa$ is also very small, the sine function is infinitely small and so the equation reduces to:

$$\frac{1}{2} g^{\lambda\nu}{}_{,0} g_{\lambda\nu,0} = -2\kappa^2 v^2 \tag{5.79}$$

All such paired metrics can use the previous technique. In so doing Einstein's tensor, with signal metric and observer moving along with the signal wave, reduces the gravitational equation to a sum of constants related directly to the local energy of spacetime. That is:

$$R_{00} = \frac{1}{2}\kappa^2 - \frac{3}{2}\kappa^2 v^2 \tag{5.80}$$

and

$$-\frac{1}{2}g^{00}R = \frac{3}{2}\kappa^2 v^2 - \frac{3}{2}\kappa^2 \tag{5.81}$$

Therefore:

$$G_{00} = \kappa^2 \eta_{00} = \frac{1}{\lambda^2}\eta_{00} \tag{5.82}$$

This result based on an exact metric $g_{\mu\nu} = [1 - \sin\kappa(x - vt)]\eta_{\mu\nu}$, matches the result from the Haar-Einstein equation.

## 6. The General Relativistic equation in Haar-wavelet Form

It was shown in the previous section that:

$$R_{00} = -\frac{3}{2}[\alpha^1_{,0}\alpha_{1,0}] + \frac{1}{2}[\alpha^1_{,1}\alpha_{1,1}] \tag{6.1}$$

and

$$-\frac{1}{2}g^{00}R = +\frac{3}{2}[\alpha^1_{,0}\alpha_{1,0}] - \frac{3}{2}[\alpha_{1,1}\alpha^1_{,1}] \tag{6.2}$$

so that:

$$G_{00} = -[\alpha^1_{,1}\alpha_{1,1}] \tag{6.3}$$

What must be shown is that the time component of the Einstein Tensor can be put into the Haar-Einstein form, expressed schematically by:

$$G\{[g^{\alpha\mu}][g_{\mu\nu}]\} \Rightarrow G\{[\tilde{\Phi}^{\kappa}{}_{\kappa}{}^{\alpha\mu}][\tilde{\Phi}^j{}_{j\mu\nu}]\} \Rightarrow \tilde{\Phi}^j{}_{j\mu\nu,\beta} =$$

$$\{a^-a_+\Phi[_{j,j+1}]\}_{\mu\nu} = [\alpha_j\alpha^j]_{\mu\nu}\Phi[_{j,j+1}] \tag{6.4}$$

with $a^-a_+$ as lowering and raising operators and $[\alpha_j\alpha^j]_{\mu\nu}$ being the eigenvalues resulting from the wave equation. To do so both $\alpha^1_{,1}, \alpha_{1,1}$ must be examined. Equation (3.2) yields $\alpha_0 = 1, \alpha_1 = \alpha^1 = \kappa(x - vt)$. Thus:

$$\alpha_{1,1} = \alpha^1_{,1} = \kappa\frac{\partial(x - vt)}{\partial x} \tag{6.5}$$

With $f'(0) = 1$, and $\eta_{00} = -1$, then:

$$\alpha_{1,1} = [\eta_{00} + f'(0)\tau]_{,x} = \kappa \tag{6.6}$$

Since $\alpha_{1,1}$ leads to a positive $\kappa$ or inverse gravitational wavelength, it is defined to be a gravitational raising operator on the metric expanded to first order. Equation (5.6) can be rewritten into the Haar wavelet term as:

$$\alpha_{1,1} = \left\{ \eta_{00}\Phi[0,\frac{1}{2}[ + f'(0)\tau\Phi[\frac{1}{2},1[ \right\}_{,x} = \kappa \tag{6.7}$$

Therefore $\alpha_{1,1}$ is the raising operator on the Haar metric.

$$\alpha_{1,1} \equiv a^+ \tag{6.8}$$

similarly, a negative $\kappa$ arises from $a_{1,0}$

$$\alpha_{1,0} \equiv a_- \tag{6.9}$$

Thus from equation (5.4) we have:

$$\tilde{\Phi}^1{}_{100,\beta} =$$
$$\left\{ a_+ \Phi_{[0,1}\cdot a_+ \Phi_{[0,1} \right\}_{00} = -[\lambda\cdot\lambda]_{00}\Phi_{[0,1[} \tag{6.10}$$

where

$$\alpha_j = \alpha^j = \lambda \tag{6.11}$$

Thus equation (5.10) is in the Haar-Einstein gravitational wave equation form:

$$\tilde{\Phi}^j{}_{j00,\beta} = -\lambda^2 \Phi[_{j,j+1} \tag{6.12}$$

## 7. Uncertainty Principle of Spacetime

Since the second term of the Haar metric can be interpreted as a graviton formed from absorption of signal energy, that is particle production from a geometric theory, the pertinent question to be addressed is: How much uncertainty arises from a measurement of energy? To answer this question, the expanded signal function is examined to first order in $\tau$:

$$g_{\mu\nu} \cong \eta_{\mu\nu} + f'(0)(x-vt)\eta_{\mu\nu} \tag{7.1}$$

The second term, $f'(0)(x-vt)\eta_{\mu\nu}$, contains signal energy, therefore signal energy is directly related to x and t, and so, too, must the uncertainty in the measurement of energy absorbed into spacetime. Fortunately, there is a well-established expression in quantum mechanics that relates energy to position and time, and is given by:

$$\Delta x \Delta p = \Delta t \Delta E \geq \frac{\hbar}{2} \tag{7.2}$$

Therefore:

$$\frac{\Delta x}{\Delta t}\Delta p = \Delta E \tag{7.3}$$

Setting $\Delta v = \frac{\Delta x}{\Delta t}$, leads to:

$$\Delta v \Delta p = \Delta v \Delta(mv) = \Delta(mv^2) = \Delta E \tag{7.4}$$

Assuming the signal wave moves at the speed of light, then:

$$\Delta E = \Delta(mc^2) \tag{7.5}$$

From this result the Theory of General Relativity to Quantum Mechanics can be strongly connected by the rule:

$$\Delta t \cdot \Delta(mc^2) \geq \frac{\hbar}{2} \tag{7.6}$$

Because the preceding equation was derived from a signal or conformal metric, and that conformal metrics are a small subset of all spacetime metrics within the scope of General Relativity, this shows that quantum mechanics is a subset of General Relativity. It also implies that probability is related to interval width. [For example, if the interval width of the Haar metric were increased from two terms to its full infinite terms, then the signal function would yield the total signal energy absorbed by spacetime. However, in so doing, little information would be gained about the particle energies emerging from spacetime. Also the relationship between relativity and the uncertainty principle would not reveal itself. In short General Relativity provides too much information because it is a continuous classical theory. Whereas, the expanded Haar metric having an infinite series of intervals and coefficients, is reduced down to two interval widths. This reduction occurs because the Taylor intervals are infinitely small, and so all higher ordered terms of the interval width can be dropped. The conclusion being, though far less information is provided by the Haar metric, which relies on interval width, it nevertheless produces decipherable information. This is precisely what wavelet theory was developed for.

## 8. Conclusion

Applying this wavelet-metric to the General Relativistic wave equation, $G_{\mu\nu}$, reduces the time component of the gravitational equation from a second ordered covariant differential equation to a first ordered partial differential equation, $\tilde{\Phi}^{j}{}_{j00,\beta}$. The Haar-Einstein equation produces the eigenvalue of $-\frac{1}{\lambda_g^2}$. Since $G_{00}$ can be approximated to the Poisson expression $\nabla^2 \phi = -4\pi\rho$, this shows the wavelength of a graviton is the fundamental source of gravitational attraction and that inertia arises from all matter in the universe as proposed by Mach and Einstein.